\documentstyle[prl,preprint,aps]{revtex}

\begin{document}

%\draft

\title{Defect--Defect Correlation Functions, Generic Scale
Invariance, and the Complex Ginzburg--Landau Equation}
\author{Bruce W. Roberts, Eberhard Bodenschatz, and James P. Sethna}
\address{Laboratory of Atomic and Solid--State Physics,\\
Cornell University,\\
Ithaca,  NY 14853--2501.}
\maketitle

\begin{abstract}
We present a calculation of defect--defect correlation functions
in the defect turbulence regime of the complex Ginzburg--Landau
equation.  Our results do not agree with the predictions of
generic scale invariance.  Using the topological nature of the
defects, we prove that defect--defect correlations
cannot decay as slowly as predicted by generic scale invariance
\end{abstract}


\narrowtext

%\section{Introduction}

Spatiotemporal chaos occurs in extended systems with many
interacting degrees of freedom\cite{CrossHo}.
Typically, it appears in nonequilibrium pattern--forming systems
slightly above their threshold of instability\cite{CrossHoRev}.  The
study of spatiotemporal chaos has been advanced by the development of
experimental systems which are precisely controlled and have a large
aspect ratio\cite{rbconvection,ehc,capillary}.
Such systems have large statistically
homogeneous regions relatively free from boundary effects.  A key
question to address is whether such regions can be described in terms
of hydrodynamic--like theories, focusing on collective behaviors and
long--wavelength descriptions.  We wish to address an aspect of this
question by considering the coherent structures known as topological
defects (or spirals or vortices) in a system that exhibits
spatiotemporal chaos: the complex Ginzburg--Landau
equation\cite{twodcgl}.  This equation
describes the slowly varying amplitude and phase in an extended system
which undergoes a Hopf bifurcation to an oscillating and
spatially uniform or oscillatory
and spatially periodic state.  The equation exhibits many interesting
patterns, but we will restrict our investigation to the Benjamin--Feir
unstable (or defect turbulent) regime\cite{Newell}, in which
topological defects occur in the context of spatiotemporal chaos.
Other systems, such as Rayleigh--B\'enard convection\cite{rbconvection},
electrohydrodynamic convection in liquid crystals\cite{ehc},
capillary ripples\cite{capillary}, cardiac tissue\cite{heart},
and chemical reactions\cite{swinney},
can exhibit similar defect--turbulent behavior.
In this letter, we will examine the
defect--defect correlation functions and relate them to the ideas of
generic scale invariance, which is a theory for describing
nonequilibrium systems with conservation laws.  We will see that
our results do not match the predictions of generic scale invariance\cite{gsi}.
Finally, we will prove that the generic predictions cannot be correct
for topological defect correlations.

%\section{The Complex Ginzburg--Landau Equation}

The complex Ginzburg--Landau equation is given by
\begin{equation}
\partial_t A = A - (1 + i c) \vert A \vert^2 A +
	( 1 + i b_x ) \frac{\partial^2 A}{\partial x^2} +
	( 1 + i b_y ) \frac{\partial^2 A}{\partial y^2}
\label{cgleqn}
\end{equation}
where $A$ is a complex field in two dimensions.
This equation
can have topological defect solutions where $A = 0$ (both ${\rm Re} A$ and
${\rm Im} A$ are zero)\cite{twodcgl,defectsolutions}.
These defects can occur in either static arrangements or in
dynamic ones called defect turbulence, where defects are
continuously nucleated and annihilated in pairs
and are moving about.  We wish to focus on the latter case,
the Benjamin--Feir turbulent instability regime\cite{Newell},
which occurs when $1 + b_\alpha c < 0$.
In this region of parameter space, all periodic solutions of the
complex Ginzburg--Landau equation are unstable.
For comparison with the
ideas of generic scale invariance\cite{gsi},
we will focus on the anisotropic case $b_x \neq b_y$.

We note that the topological
defects come in two varieties.  The type of defect depends on
how the phase of $A$ changes as we go counterclockwise once around
the defect.
A defect with a phase jump of $2\pi$ has a topological charge of $+1$,
while one with a jump of $-2\pi$ has a topological charge of $-1$.
This is analogous to the right--handed and left--handed
single--armed spirals in
Rayleigh--B\'enard convection.  We let $\rho_+({\rm \bf r})$ equal
the density of
$+1$ defects and $\rho_-({\rm \bf r})$ equal the density of $-1$ defects.
We can then define a ``topological'' order parameter,
$\rho({\rm \bf r}) \equiv \rho_+({\rm \bf r}) - \rho_-({\rm \bf r})$,
which is just the density of the
defects weighted by their topological charge.  This order parameter is
conserved in a system with periodic boundary conditions:
$\int_V \rho({\rm \bf r}) d{\rm \bf r} = 0$, as
defects can only be created or destroyed in $\pm$ pairs.
We focus on the order parameter $\rho({\rm \bf r})$ as an effective
coarse--grained field, which we conjecture can be described by
some hydrodynamic equation of motion.

%\section{Generic Scale Invariance}

In equilibrium systems, spatial correlations typically decay
exponentially.  For nonequilibrium systems (such as those
with an external driving force) the situation can be quite different.
For nonequilibrium systems with a conservation law and external noise,
spatial correlation functions can decay algebraically.  It has been
suggested that this
algebraic decay is expected to occur for a broad range of
conditions, and it has been called ``generic scale invariance''\cite{gsi}.
Some extended deterministic chaotic systems also exhibit
algebraic decay\cite{cmapl,solvedchaos,cellular}.  In at least one of these
examples the chaotic fluctuations appear to play the same role
as stochastic noise\cite{cmapl}.
The complex Ginzburg--Landau equation would seem to satisfy the
criterion for generic scale invariance.  It shows nonequilibrium
behavior since it cannot be derived from an underlying potential
({\it i.~e.} it is non--relaxational).
In a system with periodic boundary conditions,
the topological order parameter
$\rho({\rm \bf r})$ is conserved.  Finally, we conjecture that
the chaotic noise in our system plays the role of stochastic noise.

With this set of conditions, we could have a hydrodynamic equation
for the conserved order parameter of the form
\begin{equation}
\partial_t \rho({\rm \bf r},t) = \Gamma\{ \rho({\rm \bf r},t) \} +
	\eta({\rm \bf r},t)
\label{gsieqn}
\end{equation}
where $\Gamma$ is a general conserving operator on $\rho$,
such as $\Gamma_0 \nabla^2 + \Gamma_1 (\nabla^2)^2 + \Gamma_{2x}
\partial_x^4 + \Gamma_{2y} \partial_y^4$.  It can also
contain nonlinear terms ({\it e.~g.}
$\nabla \cdot [(\nabla^2 \rho) (\nabla \rho)]$ ) .
The stochastic noise term $\eta$ is determined by:
\begin{mathletters}
\label{allnoiseeqn}
\begin{equation}
\langle \eta({\rm \bf r},t) \rangle = 0
\label{noisemeaneqn}
\end{equation}
\begin{equation}
\langle \eta({\rm \bf r},t) \ \eta({\rm \bf r}~',t') \rangle =
	D \delta({\rm \bf r} - {\rm \bf r}~')\delta(t-t'),
\label{noisestddeveqn}
\end{equation}
\end{mathletters}
where $D$ must be composed of
differential operators for our strictly
conserved order parameter.
This conserving noise term represents the effect of the
chaotic fluctuations in the complex Ginzburg--Landau equation.
There is evidence from the mapping of the Kuramoto--Shivashinskii equation
to the Kardar--Parisi--Zhang equation\cite{kstokpz} and from coupled
map lattices\cite{cmapl} that this identification of spatiotemporal
chaotic fluctuations with stochastic noise is not unreasonable.

For systems with nonconserving noise ({\it i.e.} $D$ is a constant),
equation (\ref{gsieqn}) is expected to always give rise to power
law decays in the two point correlation function
$G_\rho({\rm \bf r}) \equiv \langle \rho({\rm \bf r})
\rho({\rm \bf 0}) \rangle$,
as well as in higher order correlation functions.  For systems with
conserving noise ({\it e.~g.} $D = D_1 \nabla^2$) the situation is
somewhat more complicated\cite{gsi}.  If the system is isotropic, then one
obtains exponential decays in $G_\rho({\rm \bf r})$, but power law decays
occur in higher order correlation functions.
Systems which break the isotropy give rise to algebraic decay in
$G_\rho({\rm \bf r})$.  For systems with cubic symmetry, one expects
$G_\rho({\rm \bf r}) \sim 1/r^{d+2}$ for large $r$.  For systems which
break the cubic symmetry one expects $G_\rho({\rm \bf r}) \sim 1/r^d$.
We will be working in the last regime, where in a two--dimensional
system with broken square symmetry,
generic scale invariance predicts that
\begin{equation}
G_\rho^{generic}({\rm \bf r}) =
	\langle \rho({\rm \bf r}) \rho({\rm \bf 0}) \rangle \sim 1/r^2
\label{gingsieqn}
\end{equation}
for large $r$\cite{gsi}.  We shall compare this prediction to the
results from our numerics.

Before we consider the numerics, we should provide some
caveats.  The ideas behind generic scale invariance depend upon
showing that the nonlinearities are irrelevant in a renormalization
group sense.  It has proven notoriously difficult to treat
topological defects in a perturbative manner.
An example of this is the Kosterlitz--Thouless transition\cite{kt}.
We also note that generic scale invariance requires short--ranged
interactions.  There is some evidence that this is true
for the defects in the complex Ginzburg--Landau equation\cite{Interactions},
but collective effects might be important.  Finally, the mapping
of chaotic fluctuations to stochastic noise could break down.

%\section{Numerical Results}

We have numerically solved equation (\ref{cgleqn}) with periodic
boundary conditions in the turbulent regime using a pseudo--spectral
code.  The system is $240 \times 240$ in real space, with $360$
Fourier harmonics in both directions.  We use the parameter values $c
= 1.5$, $b_x = -0.75$, and $b_y = -3.0$.  The time step used was
$\Delta t = 0.02$.  We initially ``equilibrate'' from a state with two
oppositely charged defects to a state with fluctuations about some
average number of defects.  This takes typically $5000$ time steps.
We note that the defects do not form bound pairs.  When a pair is
created, the defects tend to move apart, and when they eventually
annihilate, they usually do so with a defect other than their initial
partner.  This illustrates that we aren't in a Kosterlitz--Thouless
bound pair phase\cite{kt}.  In figure \ref{snapshotfig} we show a
snapshot of part of our system.  To find the defects in our system, we
examine the change in the phase of $A$ as we go counterclockwise
around each plaquette on our lattice in real space.  A change of $0$
signifies that the plaquette does not contain a defect, while changes
of $\pm 2 \pi$ reveal that a defect exists in the plaquette.  Once we
have found the defects, we can then proceed to calculate $n(t)$, the
total number of defects in the system, and $G_\rho({\rm \bf r})$.  To do
the averaging, we have run for $750,000$ time steps (one month of CPU
time on an IBM RS/6000 model 550: locating the defects is the time
consuming part).  We only sample $G_\rho({\rm \bf r})$ and $n(t)$ every
$10$ time steps, because adjacent time steps are not statistically
independent.  We have calculated that
$\langle n(t) n(0) \rangle - \langle n \rangle^2 \sim e^{-t/\tau}$,
with $\tau \sim 115$ time steps.
It has been predicted \cite{defectstats} that
the probability of finding a particular value of $n$ in the system is
given by $P(n) \sim e^{-(n-\langle n \rangle)^2/2\langle n \rangle}$.
We have calculated the various moments of
our distribution $P(n)$, and we find $\langle n \rangle = 422.8 \pm 0.3$,
$\sigma^2 = 397 \pm 30$, and a skewness
of $0.014$ and kurtosis of $-0.026$, which is in good agreement
with the predictions. In figure \ref{datafig}
we present the results for $G_\rho({\rm \bf r})$
with ${\rm \bf r}$ in both the
$\hat x$ and $\hat y$ directions.  For both directions the typical
nearest neighbor is of the opposite sign: the charges are thus
screened.  Similar behavior for vortices in random wave fields has
been observed\onlinecite{zeros}.  In figure \ref{loglogdatafig}
we show log--log plots of $|G_\rho(r)|$.  We also show lines that
represent the slope $|G_\rho(r)|$ should have if it decayed like
$1/r^2$.  We note that at the right edge of the figure, we have
reached the point where our data is dominated by statistical noise.
It is clear that neither direction shows the expected $1/r^2$ decay.
Our results are at variance with the predictions of generic scale
invariance.  We expect that the theory is not applicable to these
sorts of systems with strong constraints placed on them due to the
topological nature of the order parameter.  We can't explicitly
determine an equation like equation \ref{gsieqn} for the defects, but
we can discuss the results for $|G_\rho(r)|$ from the viewpoint of
topological constraints.  We will show how these constraints place
bounds upon the decay rate of the correlation function.

%\section{Topological Constraints}

We define the excess order parameter in a region to be
\begin{equation}
\delta \rho_L \equiv \left| \int_{{\rm \bf r} \in B(L)}
	\left( \rho({\rm \bf r}+ {\rm \bf r}_0) -
	\rho_0 \right) d{\rm \bf r}  \right|.
\label{deltarhoeqn}
\end{equation}
where $B(L)$ represents a circle of radius $L$ about a point ${\rm \bf r}_0$,
which we take as ${\rm \bf r}_0 = {\rm \bf 0}$ due to translational invariance.
For nontopological objects, the constraint is given
by $\delta \rho_L \leq a_1 L^2$, where $a_1$ is some numerical constant.
The excess of a nontopological object in a particular region must scale
as the volume of that region, since each individual object occupies
a fixed volume.
For topological objects, this constraint is different, {\it i.~e.}
$\delta \rho_L \leq a_2 L$, where $a_2$ is again some numerical
constant.  The constraint arises from the fact that any excess
of topological defects in a region must be detectable  simply by
traversing the perimeter of that
region.  Each topological defect has
an ``arm'' with characteristic width that
must pass through the perimeter of the region.  Examples of this
are the spiral arms of the defects in Rayleigh--B\'enard convection,
extra rows of atoms for dislocations in crystals, and in our case
lines of $Re A = 0$ and $Im A = 0$.
When a region contains the maximum excess number of defects allowed,
each of these lines takes up a fixed amount of the perimeter of the region.
Since the
maximum excess number of topological objects scales linearly with the
number of lines, and the number of lines scales as the perimeter of
the region, we must have that the maximum excess number of topological
objects scales as the linear size $L$ of the region.

If we assume that the correlation function
$G_\rho({\rm \bf r})$ decays at the same asymptotic
rate independent of the direction of ${\rm \bf r}$, {\it i.e.}
$G_\rho({\rm \bf r}) \sim f(\theta) g(r)$, where $g(r) \sim 1/r^\alpha$
for large $r$,
then with this constraint we can show for two dimensions that
$\alpha$ must be greater than 2.  This result also requires that
$\int_0^{2 \pi} f(\theta) d\theta \neq 0$, which
we expect to be true except for special cases.
An example of such a correlation
function satisfying both assumptions is given in
reference \onlinecite{kawasaki}.
To show that $\alpha > 2$, we begin by
squaring equation (\ref{deltarhoeqn}) and
averaging the result (over the noise or over time and space).
This gives us the constraint equation
\begin{equation}
\langle\delta \rho_L^2\rangle = \int_{{\rm \bf r} \in B(L)}
	\int_{{\rm \bf r}~' \in B(L)} d{\rm \bf r} d{{\rm \bf r}~'}
	G_\rho({\rm \bf r} - {\rm \bf r}~') \leq a^2 L^2.
\label{avgconstrainteqn}
\end{equation}
Separating $G_\rho({\rm \bf r} - {\rm \bf r}~')$ into radial
and angular components yields
\begin{equation}
\langle\delta \rho_L^2\rangle = \frac{1}{2 \pi}
	\int_0^{2\pi} f(\theta) d\theta \int_0^{2L} g(R) w(R) dR
\label{splitconstrainteqn}
\end{equation}
where we can derive that
\begin{equation}
w(R) = 4 \pi R L^2 \left[ \cos^{-1}\left(\frac{R}{2L}\right) -
	\frac{R}{2L} \sqrt{1-\left(\frac{R}{2L}\right)^2} \right].
\label{wreqn}
\end{equation}
This result is obtained by considering how often
$|{\rm \bf r} - {\rm \bf r}~'| = R$ when ${\rm \bf r} \in B(L)$ and
${\rm \bf r}~' \in B(L)$, {\it i.~e.}
\begin{equation}
w(R) = \int_{{\rm \bf r} \in B(L)} \int_{{\rm \bf r}~' \in B(L)}
	d{\rm \bf r} d{{\rm \bf r}~'} S(R)
	\delta({\rm \bf r} - {\rm \bf r}~' - {\rm \bf R}),
\label{wrdefeqn}
\end{equation}
where $S(R) = 2\pi R$ (the circumference of the circle of radius $R$).
{}From this definition $w(R)$ can be calculated by writing the
$\delta$--function in integral form, and then performing the
resulting integrals.
By assuming that $G_\rho({\rm \bf R})$ exhibits its asymptotic behavior
outside of some $R > r_{min}$, we can split
the radial integral in equation (\ref{splitconstrainteqn}) into two parts:
\begin{equation}
\int_0^{2 L} g(R) w(R) dR = \int_{r_{min}}^{2 L} + \int_0^{r_{min}}
	g(R) w(R) dR.
\label{radialspliteqn}
\end{equation}
We can then examine the large $L$ limit of
$\langle \delta \rho_L^2\rangle$
for various values of the power law exponent $\alpha$.  By rescaling
all lengths by $2L$ and then expanding $w(R)$ about small $R$, we
can show that the leading order behaviour of $\langle \delta \rho_L^2\rangle$
is
\begin{eqnarray}
\alpha < 2 & : & \langle \delta \rho_L^2 \rangle
	\sim L^2 L^{2-\alpha} \nonumber \\
\alpha = 2 & : & \langle \delta \rho_L^2 \rangle
	\sim L^2 \log(L) \nonumber \\
\alpha > 2 & : & \langle \delta \rho_L^2 \rangle
	\sim L^2.
\label{alphaeqn}
\end{eqnarray}

Within our assumptions, this result means that for topological objects
the results of generic scale invariance cannot hold, since the
constraint given by equation (\ref{avgconstrainteqn}) would be
violated.
In fact, what we have provided is a bound on $\alpha$.  For topological
objects $\alpha$ must be strictly greater than $2$.
Generic scale invariance predicts $\alpha = 2$.
The simple geometric nature of topological objects
prevents them from having correlation
functions that decay as certain power laws.  An added conclusion is that if
the topological objects form ordered states, they must be
of the antiferromagentic variety ({\it e.g.} alternating $+$ and $-$
vortices) in at least one direction,
in order to satisfy the topological constraint.  An example of such a
state is given in \cite{advmat}.

In our analysis we have only considered the largest possible fluctuations.
We expect
these fluctuations to be rare, and hence expect a faster decay
than the bound we provide.  As an analogy, for nontopological
objects the analysis presented here would predict
that the correlation function can be at most a
constant for large r; in practice systems like spins or atoms have
connected correlation functions that
decay to zero, either as
power laws or as exponentials.
We expect similarly for real topological objects the decay of the
correlation functions will be faster than the bound given here.
Numerically, we have found that we are indeed far from
saturating this bound.

%\section{Conclusions}

%\section{Acknowledgements}

We would like to thank P.~C.~Hohenberg, J.~F.~Marko, and M.~E.~J.~Newman
for helpful conversations.  This work was
partly funded by the Hertz Foundation~(BWR), the NSF under grant
DMR--91--18065~(BWR,~JPS), and the Alfred P. Sloan Foundation (EB).
We also thank the Cornell Materials Science Center for the use
of its computational resources.

% References
%

% Figures
%
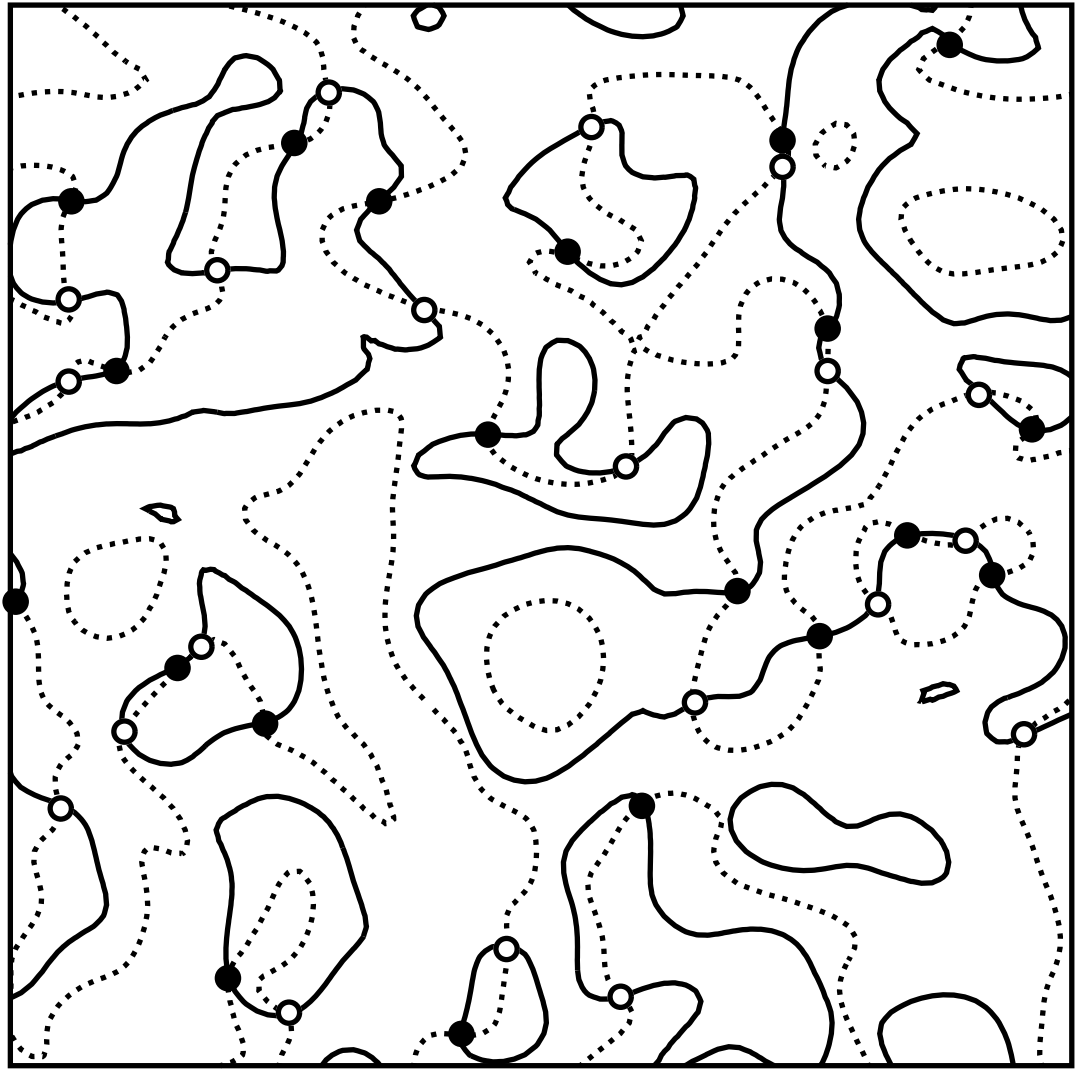
\begin{figure}
\caption{Snapshot of a $70 \times 70$ region. The solid
lines are where $Re A=0$, and the dashed lines are where $Im A = 0$.
Filled circles ($\bullet$) are vortices with
topological charge $+1$, and the open
circles ($\circ$) have charge $-1$.}
\label{snapshotfig}
\end{figure}
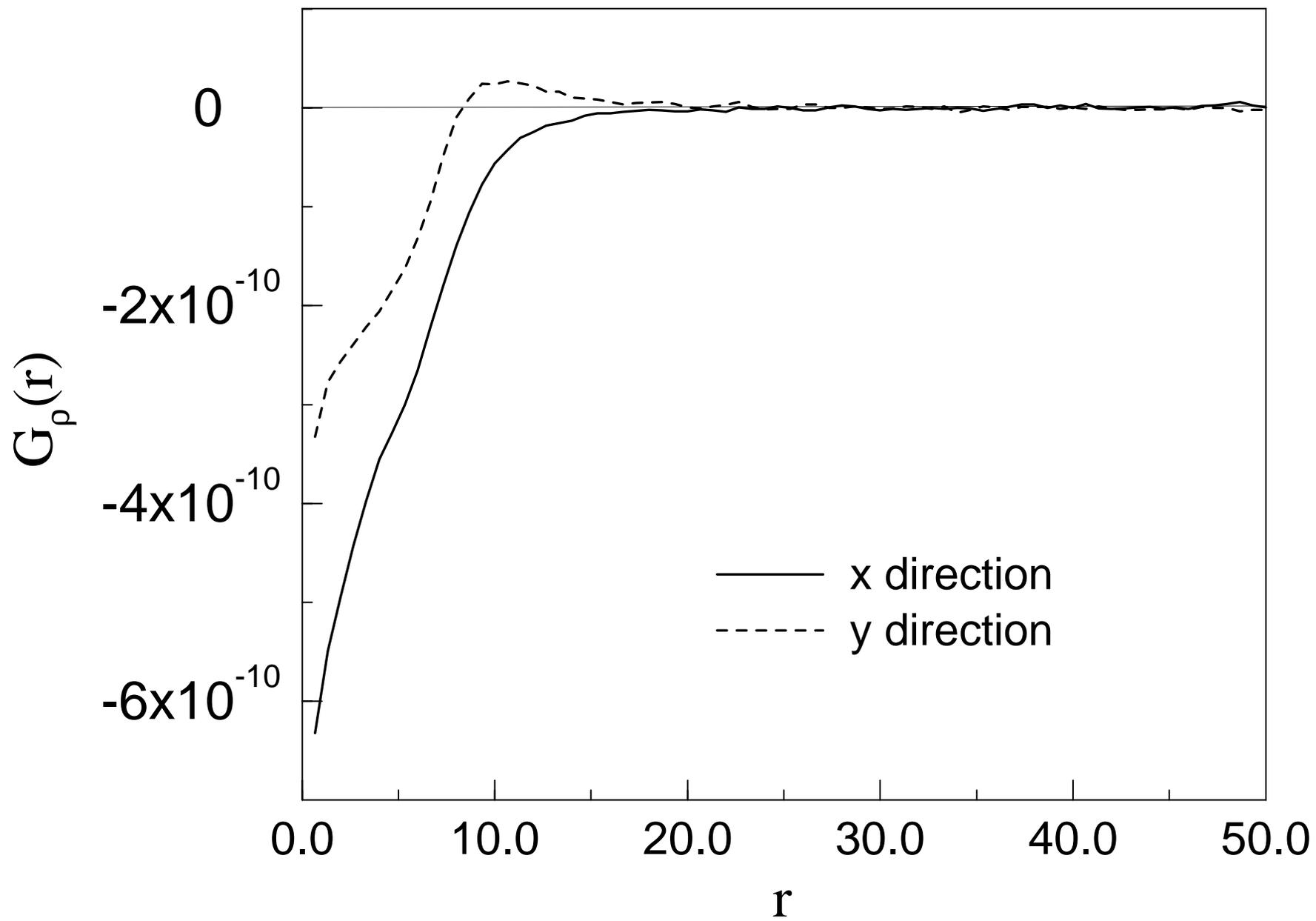
\begin{figure}
\caption{$G_\rho({\rm \bf r})$ versus $r$.  The solid line is for the
	${\hat x}$ direction, while the dashed line is for the ${\hat y}$
	direction.  Also shown is a line for $G=0$.  Note that
	$G$ attains its asymptotic limit of $0$ from different
	sides of this line.}
\label{datafig}
\end{figure}
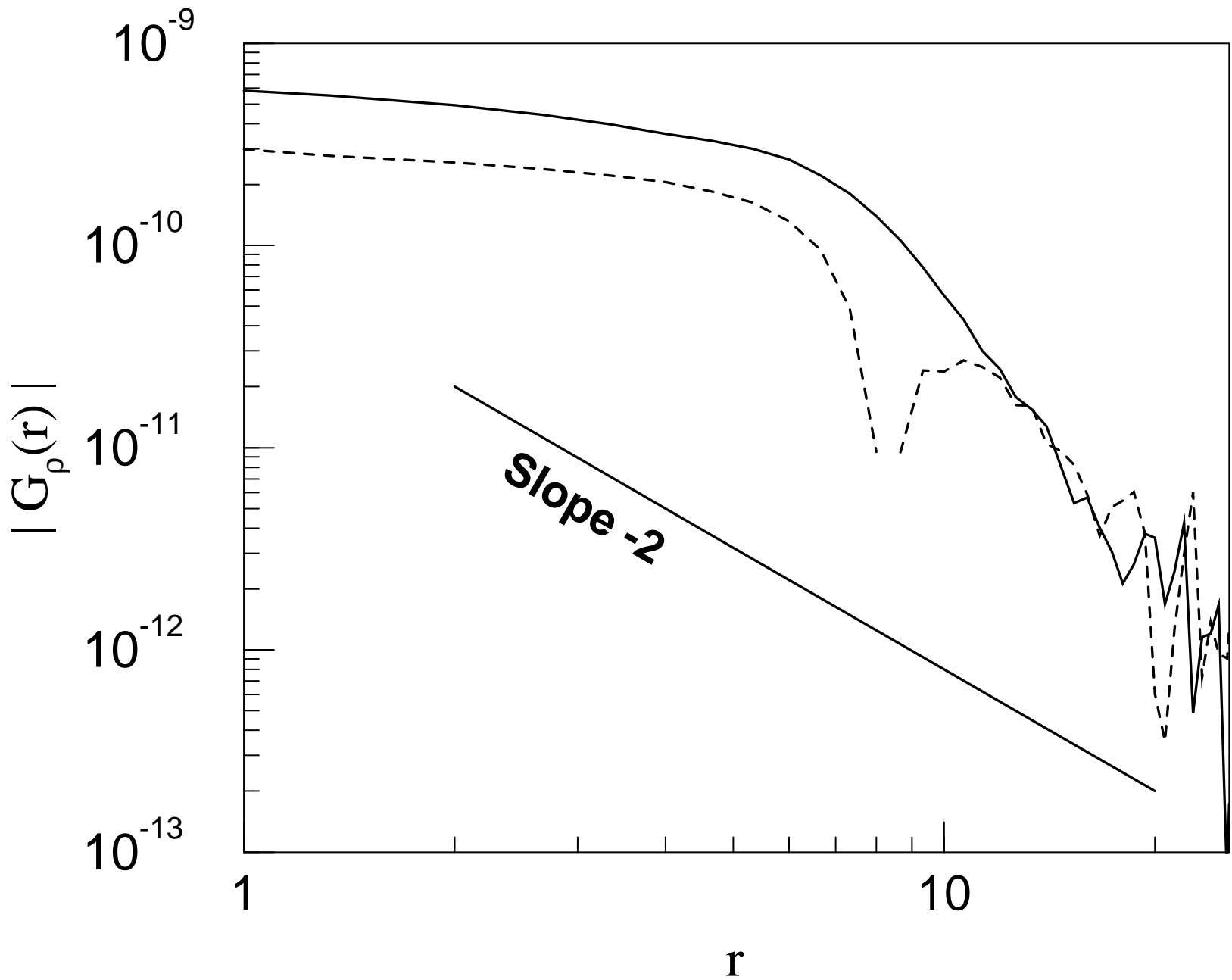
\begin{figure}
\caption{Log--log plot of $| G_\rho({\rm \bf r}) |$ versus $r$.
	The solid line corresponds to the ${\hat x}$ direction and
	the dashed line to the ${\hat y}$ direction.
	Also shown is a line with slope that would correspond to
	$|G_\rho(r)| \sim 1/r^2$. Note also the break in
	the line for the ${\hat y}$ direction which corresponds
	to the zero crossing of $G$.}
\label{loglogdatafig}
\end{figure}

\begin{references}
%
\bibitem{CrossHo} M. C. Cross and P. C. Hohenberg, Science {\bf 263},
	1569 (1994).
%
\bibitem{CrossHoRev} M. C. Cross and P. C. Hohenberg,
	Reviews of Modern Physics {\bf 65}, 851 (1993).
%
\bibitem{rbconvection} E. Bodenschatz, J. DeBruyn, G. Ahlers, and
	D. Cannell, Phys. Rev. Lett. {\bf 67}, 3078 (1991);
	Stephen W. Morris, Eberhard Bodenschatz,
	David S. Cannell, and Guenter Ahlers, Phys. Rev. Lett.
	{\bf 71}, 2026 (1993).
%
\bibitem{ehc} S. Kai and K. Hirakawa, Prog. Theor. Phys. Suppl.,
	{\bf 64}, 212 (1978);
	E. Braun, S. Rasenat, and V. Steinberg, Europhys. Lett. {\bf 15},
	597 (1991).
%
\bibitem{capillary} N. B. Tufillaro, R. Ramashankar, and J. P. Gollub,
	Phys. Rev. Lett. {\bf 62}, 422 (1989).
%
\bibitem{twodcgl} P. Coullet, L. Gil, and J. Lega, Phys. Rev. Lett.
	{\bf 62}, 1619 (1989); Physica D {\bf 37}, 91 (1989);
	Igor Aranson, Lorenz Kramer, and Andreas Weber, Phys. Rev. Lett.
	{\bf 72}, 2316 (1994).
%
\bibitem{Newell} A. C. Newell, in {\it Nonlinear Wave Motion}, edited
	by A. C. Newell (American Mathematical Society, Providence,
	R. I., 1974).
%
\bibitem{heart} J. M. Davidenko {\it et al.}, Nature {\bf 355}, 349
	(1993).
%
\bibitem{swinney} Q. Ouyang and H. L. Swinney, Chaos {\bf 1}, 411 (1991).
%
\bibitem{gsi} G. Grinstein, D.--H. Lee, and Subdir Sachdev, Phys. Rev. Lett.
	{\bf 64}, 1927 (1990); G. Grinstein, J. Appl. Phys. {\bf 69}, 5441
	(1991); G. Grinstein, C. Jayaprakash, and J. E. S. Socolar,
	Phys. Rev. E {\bf 48}, R643 (1993).
%
\bibitem{defectsolutions} P. S. Hagan, SIAM J. Appl. Math. {\bf 42},
	762 (1982); E. Bodenschatz, A. Weber, and L. Kramer,
	in {\it Nonlinear Wave Processes in Excitable Media},
	A. Holden, M Markus, and H. Othmer, eds. (Plenum Press,
	New York, 1991).
%
\bibitem{cmapl} Ravi Bhagavatula, G. Grinstein, Yu He, and C.
	Jayaprakash, Phys. Rev. Lett. {\bf 69}, 3483 (1992).
%
\bibitem{solvedchaos} D. Hansel and H. Sompolinsky, Phys. Rev. Lett.
	{\bf 71}, 2710 (1993).
%
\bibitem{cellular} M. Eisele, Physica D {\bf 48}, 295 (1991).
%
\bibitem{kstokpz} C. Jayaprakash, F. Hayot, and Rahul Pandit,
	Phys. Rev. Lett. {\bf 71}, 12 (1993).
%
\bibitem{kt} P. Minnhagen, Rev. Mod. Phys. {\bf 59}, 1001 (1987).
%
\bibitem{Interactions} Igor S. Aranson, Lorenz Kramer, and Andreas Weber,
	Phys. Rev. E {\bf 47}, 3231 (1993).
%
\bibitem{defectstats} L. Gil, J. Lega, J. L. Meunier, Phys. Rev. A
	{\bf 41}, 1138 (1990).
%
\bibitem{zeros} N. Shvartsman and I. Freund, Phys. Rev. Lett. {\bf 72},
	1008 (1994).
%
\bibitem{kawasaki} Z. Cheng, P. L. Garrido, J. L. Lebowitz, and
	J. L. Vall\'es, Europhys. Lett. {\bf 14}, 507 (1991).
%
\bibitem{advmat} A. Weber, E. Bodenschatz, and L. Kramer,
	Adv. Mater. {\bf 3}, 191 (1991).
%
\end{references}
\end{document}